# Real-Time, Crowdsourcing-Enhanced Forecasting of Building Functionality During Urban Floods

Lei Xie[1], Peihui Lin[1], Naiyu Wang[1✉], Paolo Gardoni[2]

[1] College of Civil Engineering and Architecture, Zhejiang University, Hangzhou, Zhejiang 310058, China.
[2] University of Illinois Urbana-Champaign, IL, USA
✉Correspondence: naiyuwang@zju.edu.cn

**ABSTRACT:** Urban flood emergency response increasingly relies on infrastructure impact forecasts rather than hazard variables alone. However, real-time predictions are unreliable due to biased rainfall, incomplete flood knowledge, and sparse observations. Conventional open-loop forecasting propagates impacts without adjusting the system state, causing errors during critical decisions. This study presents CRAF (Crowdsourcing-Enhanced Real-Time Awareness and Forecasting), a physics-informed, closed-loop framework that converts sparse human-sensed evidence into rolling, decision-grade impact forecasts. By coupling physics-based simulation learning with crowdsourced observations, CRAF infers system conditions from incomplete data and propagates them forward to produce multi-step, real-time predictions of zone-level building functionality loss without online retraining. This closed-loop design supports continuous state correction and forward prediction under weakly structured data with low-latency operation. Offline evaluation demonstrates stable generalization across diverse storm scenarios. In operational deployment during Typhoon Haikui (2023) in Fuzhou, China, CRAF reduces 1–3 hour-ahead forecast errors by 84–95% relative to fixed rainfall-driven forecasting and by 73–80% relative to updated rainfall-driven forecasting, while limiting computation to 10 minutes per update cycle. These results show that impact-state alignment—rather than hazard refinement alone—is essential for reliable real-time decision support, providing a pathway toward operational digital twins for resilient urban infrastructure systems.

**KEYWORDS:** Building functionality; Crowdsourcing; Deep-learning; Impact-based forecasting; Rolling updates; Spatiotemporal forecasting; Urban floods

## 1 INTRODUCTION

Urban flooding, intensified by climate change and rapid urbanization, is occurring with increasing frequency and severity worldwide, posing escalating risks to public safety, infrastructure performance, and the continuity of essential services. Recent events highlight the magnitude of this challenge: the 2012 Beijing rainstorm caused 79 fatalities and RMB 11.64 billion in economic losses (Zhu et al. 2022), while the 2021 floods in Germany resulted in 189 fatalities and damages exceeding €40 billion (Meyer and Johann 2025; Zander et al. 2023). Beyond these direct losses, such events reveal systemic vulnerabilities in how urban infrastructure systems are monitored, assessed, and managed during rapidly evolving hazards.

While advances in meteorological and hydrological modeling have improved the prediction of hazard variables such as rainfall and inundation, emergency decisions are rarely made on hazards alone. Instead, operational actions—such as evacuation, shelter activation, and resource allocation—depend on reliable estimates of impacts, including loss of building functionality, accessibility disruptions, and service interruptions. From a systems and reliability perspective, the critical challenge is therefore not only to forecast hazards, but to estimate and continuously update the evolving impact state of the built environment in real time. Translating uncertain and rapidly changing hazards into reliable, rolling decision-relevant impact forecasts remains a central yet unresolved problem in flood risk management.

In practice, real-time impact forecasting is hindered by multiple sources of uncertainty. Precipitation forecasts during extreme events often exhibit substantial bias and variability, which propagate through hydrologic and hydraulic models and degrade downstream consequence predictions (Mazzoleni et al. 2017; Guan et al. 2023; Songchon et al. 2023). Flood-conditioning factors—such as drainage capacity, surface roughness, and antecedent conditions—are imperfectly observed and difficult to calibrate during unfolding events (Su et al. 2025). Meanwhile, conventional monitoring networks are designed primarily for hazard variables and lack the spatial density and temporal responsiveness required to capture the heterogeneous progression of impacts across urban communities (Farahmand et al. 2023; Lee and Tien 2018; Wang et al. 2018a; Wang et al. 2018b; Panakkal and Padgett 2024). As a result, impact forecasts frequently diverge from reality precisely during the narrow time windows when timely and reliable situational awareness is most critical.

Crowdsourced observations—including social media posts, citizen reports, and online news—provide a complementary source of real-time information that is abundant, low-cost, and often available with minimal latency during disaster evolution (Fohringer et al. 2015; Rossi et al. 2018; Songchon et al. 2023). These human-sensed data streams can reveal localized evidence of flooding and disruption that is not captured by traditional sensors (Table 1). Yet effectively translating these heterogeneous reports into operational impact forecasting remains challenging. Crowdsourced data are inherently unstructured (text, images, and video), unevenly distributed in space and time, and subject to reporting noise and ambiguity. These characteristics limit the effectiveness of classical data assimilation methods (e.g., Kalman-filter variants) that rely on structured measurements with well-characterized uncertainties (Mazzoleni et al., 2017; Annis & Nardi, 2019; Songchon et al., 2023), as well as purely data-driven regression models that require dense, consistently labeled training data (Yuan et al., 2023; Safaei-Moghadam et al., 2024). Consequently, most existing approaches employ crowdsourcing to refine hazard estimations, rather than to directly update system-level impact state forecasts that are most relevant to real-time decision-making (Assumpção et al. 2021).

Table 1. Comparison of crowdsourcing-informed flood forecasting studies with CRAF

| Study | Domain | Physics-Trained Prior (Offline) | Crowd Assimilation (Online) | Target= Functionality/ Impact | Methodology |
|---|---|---|---|---|---|
| Mazzoleni et al., 2017 | Flood | ✗ | ✓ | ✗ (streamflow) | Semi-Distributed Hydrological Model + Kalman filter |
| Annis & Nardi, 2019 | Flood | ✗ | ✓ | ✗ (water depth/extent) | FLO-2D PRO + Ensemble Kalman filter |
| Restrepo-Estrada et al., 2018 | Flood | ✗ | ✓ | ✗ (streamflow) | Social media → Rainfall proxy → Probability Distributed Model |
| Songchon et al., 2023 | Flood | ✗ | ✓ | ✗ (water depth/extent) | LISFLOOD-FP + Ensemble Kalman filter |
| Yuan et al., 2023 | Road flooding | ✗ | ✗ | ✓ (road flooding risk) | ML model (RF/AdaBoost) trained on crowd labels |
| Safaei-Moghadam et al., 2024 | Road flooding | ✗ | ✗ | ✓ (road flooding risk) | ML model (RF/XGBoost/SVC) trained on flood alerts |
| This work (CRAF) | Building functionality | ✓ (encodes inter-ERZ correlation) | ✓ (uses prior to fuse crowd) | ✓ (building functionality loss) | Graph-based situation awareness + spatiotemporal forecasting (physics-supervised) |

**Notes**：✓ = present; ✗ = absent. *Physics-trained prior* encodes inter-zone impact correlations learned from large-scale simulations for online inference (beyond simply including physics-inspired features). *Crowd assimilation* incorporates real-time human-sensed observations to update forecasts. *Functionality/impact* denotes operationally meaningful outcomes (e.g., building functionality, road operability) rather than hydraulic variables.

These limitations suggest that reliable real-time impact forecasting requires more than improved hazard prediction or isolated observations. Instead, it calls for a **closed-loop, system-level framework** in which impacts are treated as a latent state of the infrastructure network that is repeatedly estimated, corrected, and propagated as new evidence becomes available. Such a formulation enables continuous alignment between model predictions and evolving field conditions, thereby enhancing the reliability of forecasts used for operational decisions. In contrast, conventional open-loop approaches propagate impacts forward based solely on forcing inputs, allowing errors to accumulate and reducing their usefulness during critical response periods.

Motivated by this need, this study introduces CRAF (Crowdsourcing-Enhanced Real-Time Awareness and Forecasting), a physics-informed framework that performs closed-loop impact-state alignment and forecasting, transforming sparse human-sensed evidence into rolling, decision-grade impact predictions (Fig. 1). Using zone-level building functionality loss (ZFL) as a representative, decision-relevant metric, CRAF formulates real-time impact forecasting as a sequential state estimation and propagation problem, in which the evolving system state is continuously reconstructed and updated as new observations emerge. The framework operates through three tightly coupled functional streams. Crowdsourced Impact Monitoring (CIM) extracts sparse, time-stamped impact cues from heterogeneous online sources. Situational Awareness (SA) assimilates these observations through a physics-trained prior to infer spatially coherent impact states across all zones, even under severe observation sparsity. Spatiotemporal Forecasting (STF) then propagates the calibrated state forward under rainfall forcing to generate rolling, multi-step forecasts. By separating offline physics-supervised learning from online low-latency inference, CRAF enables repeated state correction and forecasting without retraining, thereby providing a stable, operationally viable, and data-efficient closed-loop capability for real-time impact forecasting.

This study advances real-time flood impact forecasting through three complementary developments. First, it introduces a closed-loop impact-state forecasting paradigm that continuously aligns predictions with evolving field evidence, improving reliability under uncertainty. Second, it develops a physics-informed forecasting framework that integrates crowdsourced observations with simulation-based knowledge to enable robust state estimation when data are sparse, heterogeneous, and incomplete. Third, the framework is demonstrated through operational deployment during a real flood event, demonstrating substantial reductions in short-horizon forecast errors during decision-critical periods. Together, these elements establish a practical foundation for real-time, impact-based decision support and more resilient urban infrastructure systems.

The remainder of this paper is organized as follows. Section 2 presents the problem formulation and defines the impact variable and spatial units. Section 3 describes the datasets, physics-based simulations, and preprocessing procedures. Section 4 details the design and training of the CRAF modules. Section 5 evaluates operational performance through a real-world case study. Section 6 concludes with key findings and future research directions.

## 2 PROBLEM FORMULATION AND FRAMEWORK

### 2.1 Spatial Unit and Predictive Impact Variable

To operationalize flood-induced impacts for emergency management, we focus on building functionality loss as the target impact variable. Building functionality represents the ability of the building to maintain structurally safe occupancy, provide the intended services to the tenants (e.g., potable water and power), and have physical access under flooding, accounting for inundation, physical damage, utility disruptions, and accessibility constraints (Chavez et al., 2025; Lin & Wang, 2017; Nocera & Gardoni, 2019; Sun & Cha, 2022). Loss of functionality directly informs evacuation and shelter decisions and therefore serves as a decision-relevant indicator for emergency response (Xie et al. 2025b; Guidotti et al. 2019; Lamadrid et al. 2025).

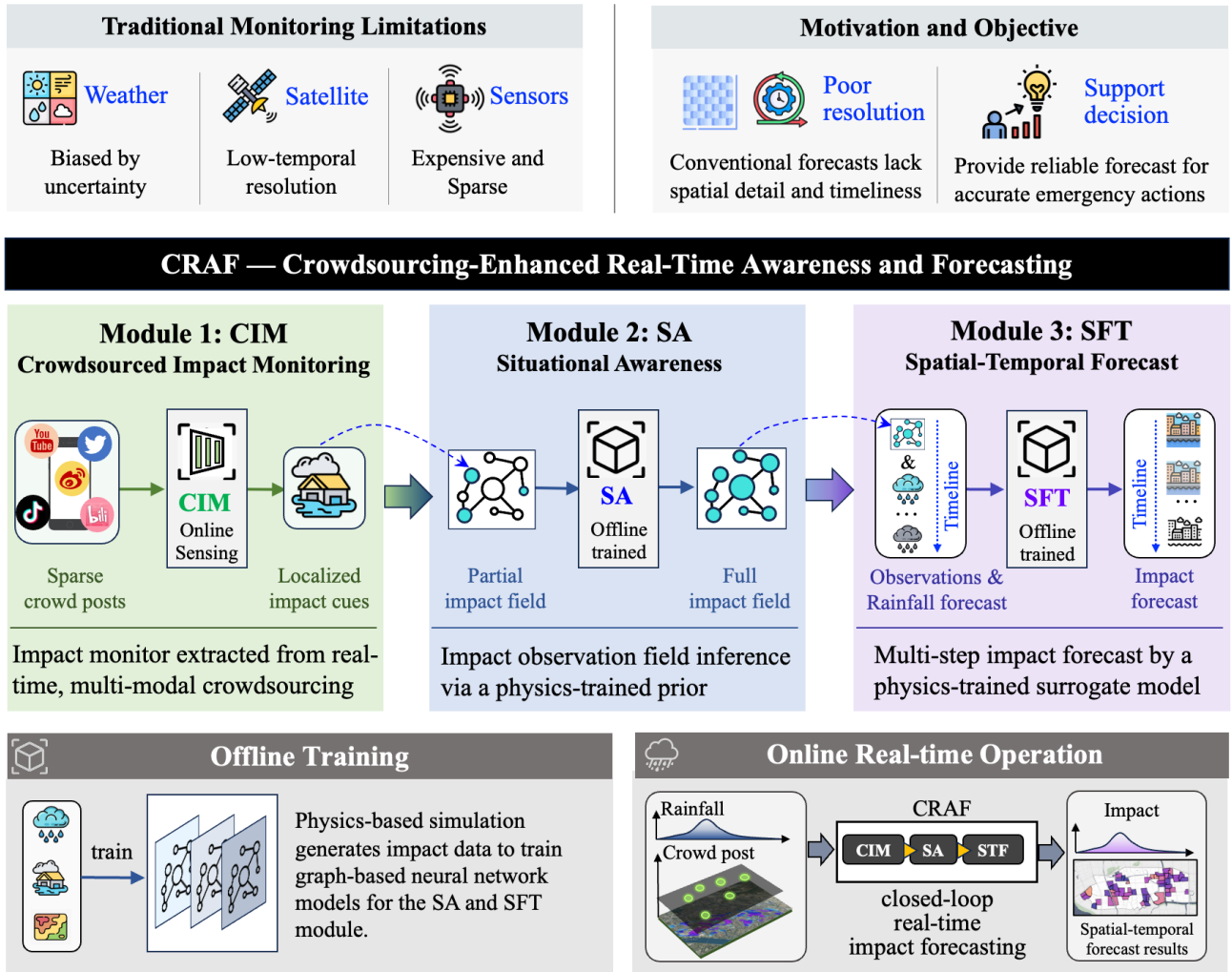

Fig. 1 Conceptual overview of the CRAF framework.

Emergency actions are typically coordinated at aggregated spatial units rather than at the level of individual buildings. Accordingly, the study area is partitioned into Emergency Response Zones (ERZs), each comprising a cluster of geographically proximate buildings with similar flood-conditioning characteristics. Let the ERZs be indexed by $\mathcal{M} = \{1, ..., M\}$, where $M$ is the total number of ERZs. For ERZ $m \in \mathcal{M}$ containing $N_m$ buildings, the zone-level functionality loss (ZFL) at time $t$, denoted by $z_m^t \in [0,1]$, is defined as the average probability that buildings within the zone are in non-functional states requiring evacuation or external assistance:

$$z_m^t = \frac{1}{N_m} \sum_{i=1}^{N_m} \sum_{F_k \in \mathcal{L}} p_{i,t}^{F_k} \in [0,1] \quad (1)$$

Here $p_{i,t}^{F_k}$ is the probability that building $i$ is in functionality state $F_k$ at time $t$, and $\mathcal{L}$ denotes the set of non-functional states that mandate evacuation action (as detailed in Section 3.3).

Vectorizing across all ERZs yields the system-level impact state at time $t$:

$$Z_t = (z_1^t, z_2^t, ..., z_M^t) \in [0,1]^M \quad (2)$$

This vector-valued representation defines the fundamental impact state to be sequentially estimated, updated, and propagated throughout the framework.

### 2.2 Impact Forecasting Problem Description

The objective of this paper is to forecast the spatiotemporal evolution of flood-induced impact states, represented by ZFL at ERZ resolution, in real time, accounting for errors and incompleteness in meteorological forcing, flood-conditioning attributes, and observational evidence.

Let $H$ denote the look-back window and $F$ the forecast horizon. At each forecast initialization time $t$, the framework ingests three categories of inputs:

- **Rainfall forcing.** Observed rainfall over the look-back window, $R_{t-H+1:t}$, and forecast rainfall over the prediction horizon, $R_{t+1:t+F}$, concatenated into a unified rainfall sequence $R_{t-H+1:t+F}$.
- **Static ERZ-level attributes.** Flood-conditioning features $W \in \mathbb{R}^{M \times d_w}$, where $d_w$ denotes the number of flood-conditioning features capturing spatial heterogeneity in flood susceptibility, and ERZ-level building attributes $E \in \mathbb{R}^{M \times d_e}$, where $d_e$ denotes the number of building attributes describing the composition of the built environment.
- **Crowdsourced observations.** A collection of crowdsourced posts $P_{\mathcal{T}_t} = \{P_\tau \mid \tau \in \mathcal{T}_t\}$, where $\mathcal{T}_t \subseteq [t-H+1, t]$ denotes the set of time indices within the look-back window where crowd observations are available.

The forecasting task is to map these heterogeneous inputs into a multi-step forecast trajectory of the impact state:

$$\mathcal{F}: (R_{t-H+1:t+F}, W, E, P_{\mathcal{T}_t}) \to \hat{Z}_{t+1:t+F} \quad (3)$$

This formulation highlights two defining characteristics of real-time impact forecasting: (i) the current impact state is latent and must be inferred from sparse, noisy observations, and (ii) forecast trajectories must remain robust to rainfall-forcing uncertainty while supporting low-latency updates. Impact forecasting is therefore naturally viewed as a sequential state estimation and propagation process, rather than a one-shot direct prediction of future impacts. Under this formulation, open-loop methods propagate impact states using rainfall forcing alone, whereas closed-loop methods incorporate recurrent state correction when new impact observations become available.

## 2.3 Overview of the CRAF Framework

CRAF is designed as an integrated three-module impact-state estimation and forecasting system:

**CIM Module ($f_{CIM}$) — observation operator.** From the collection of raw crowdsourced posts $P_{\mathcal{T}_t} = \{P_\tau \mid \tau \in \mathcal{T}_t\}$, the CIM module extracts time- and location-specific inundation cues from heterogeneous sources (text, images, or video) and converts them into sparse ERZ-level impact observations $O_{\mathcal{T}_t} = \{O_\tau \mid \tau \in \mathcal{T}_t\}$. At each time $\tau$, a single observation is represented as $O_\tau = \{(m, z_m^\tau) \mid m \in O_\tau^*\}$, where $O_\tau^* \subset \mathcal{M}$ denotes the subset of ERZs for which reliable crowd evidence is available. The observation operator is defined as:

$$O_\tau = f_{CIM}(P_\tau), \quad \tau \in \mathcal{T}_t \quad (4)$$

CIM does not attempt to produce spatially complete impact fields; instead, it provides sparse but informative measurements of the underlying impact state. The CIM module produces **deterministic** ERZ-level impact observations and does not explicitly model observation noise or uncertainty. Instead, potential errors in the extracted observations are mitigated through a **strict quality-control process**, allowing the resulting observations to be treated as **high-confidence inputs** in the subsequent data assimilation framework.

**SA Module ($f_{SA}$) — spatial impact state completion.** The SA module transforms sparse observations into a spatially complete and physically coherent impact state estimate. Given observations $O_\tau$ and static ERZ attributes $(W, E)$, the SA module infers a dense ZFL snapshot

$$Y_\tau = f_{SA}(O_\tau, W, E; G_1) \in [0,1]^M \quad (5)$$

where $G_1 = (V, A_1)$ is an ERZ graph whose adjacency matrix encodes inter-zone impact co-variation learned from physics-based simulations. For observed zones $m \in O_\tau^*$, the inferred state satisfies $(Y_\tau)_m = z_m^\tau$; for unobserved zones, values are inferred through graph-based propagation. Its parameters are learned offline from physics-generated ZFL trajectories under simulated sparse-observation conditions.

**STF Module ($f_{STF}$) — spatiotemporal impact state propagation.** The STF module propagates calibrated impact states forward in time under rainfall forcing. Given rainfall inputs $R_{t-H+1:t+F}$ and a history of impact states composed of inferred states $Y_{\mathcal{T}_t}$ and predicted states at unobserved times, the STF module produces rolling multi-step forecasts:

$$\hat{Z}_{t+1:t+F} = f_{STF}(R_{t-H+1:t+F}, [\hat{Z}_{\mathcal{T}_t^c} \| Y_{\mathcal{T}_t}]; G_2) \quad (6)$$

where $\mathcal{T}_t^c \subseteq [t - H + 1, t]$ denotes time steps without crowd observations, ∥ denotes temporal concatenation, and $G_2 = (V, A_2)$ is a graph encoding correlations among ERZ-level ZFL temporal trajectories. The STF module is trained offline using physics-generated ZFL sequences and corresponding rainfall forcing, ensuring consistency between spatial inference and temporal propagation.

## 2.4 Offline Physics-Supervised Learning and Online Forecast Operation

CRAF integrates offline physics-supervised learning with online impact-state inference to enable rolling real-time forecasts. Prior to deployment, the SA and STF modules are trained offline using large ensembles of physics-based flood and building-functionality simulations. These simulations generate spatiotemporally correlated ZFL trajectories under diverse rainfall and flood-conditioning scenarios, allowing the models to learn (i) persistent spatial co-variation of impacts across ERZs and (ii) temporal evolution patterns of functionality loss conditioned on rainfall forcing. Once trained, the parameters of both modules are fixed and remain unchanged during real events.

During a flood event, CRAF operates entirely in online inference mode (Fig. 2). The CIM module continuously ingests heterogeneous public posts and converts them into sparse ERZ-level impact observations. These observations are assimilated by the pre-trained SA module to infer a spatially complete estimate of the current impact state, which is then propagated forward by the pre-trained STF module to generate rolling multi-step forecasts. As new crowd evidence becomes available, impact states and forecasts are updated sequentially without retraining. This repeated CIM→SA→STF cycle closes the loop between sensing, state estimation, and rainfall-conditioned propagation, enabling continuous realignment of the latent impact state with evolving real-world conditions while maintaining low-latency operation.

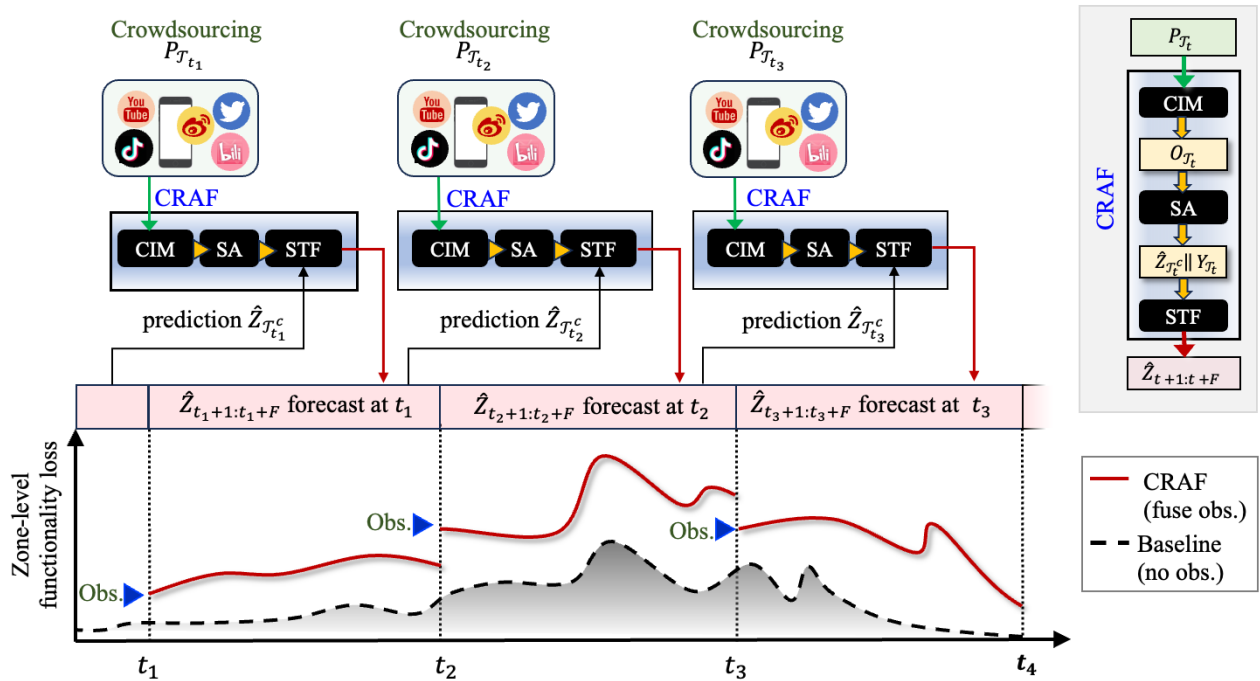

Fig. 2 The application flowchart of CRAF framework for dynamic forecast updates.

## 3 DATA DESCRIPTION AND PREPARATION FOR MODEL TRAINING

This section describes the data required to instantiate and train the CRAF framework. Consistent with its system-oriented design, CRAF relies on a limited set of broadly available data categories rather than location-specific or bespoke inputs. These data support representation of static exposure and flood-conditioning characteristics, dynamic meteorological forcing, physics-based supervisory labels for offline learning, and real-time observational evidence for online impact-state updating.

To illustrate the practical implementation, the framework is instantiated for a representative flood-prone urban community in Fuzhou, China. While specific datasets are drawn from this case study, the framework itself is not tied to any particular city, sensor network, or simulation platform.

### 3.1 General Data Requirements of the CRAF Framework

Application of the CRAF framework requires four categories of data, corresponding directly to the modeling components introduced in Section 2.

**(1) ERZ-level building attributes**

Building and infrastructure data are required to characterize the composition of the built environment within each ERZ. At a minimum, these attributes include building typology or archetype proportions, which enable aggregation of building-level impacts into ZFL. These attributes are treated as static features and represented as an ERZ-level matrix $E \in \mathbb{R}^{M \times d_e}$.

**(2) Meteorological forcing**

Time-varying meteorological inputs—primarily rainfall—drive impact-state propagation in the forecasting module. Rainfall is represented as hourly sequences spanning a historical look-back window $H$ and a forecast horizon $F$, denoted as $R \in \mathbb{R}^{(H+F) \times N_R}$, where $H + F$ is the sequence length and $N_R$ is the number of rainfall inputs. Data may originate from gauges, radar products, or numerical weather prediction outputs.

**(3) Flood-conditioning variables**

Flood-conditioning variables capture static spatial heterogeneity governing flood accumulation and impact patterns, including topography, drainage characteristics, and proximity to waterways. These variables are aggregated at the ERZ level to form a conditioning matrix $W \in \mathbb{R}^{M \times d_w}$.

**(4) Physics-based simulation outputs**

Offline training of the SA and STF modules relies on physics-based flood and impact simulations to generate supervisory labels. These simulations provide internally consistent spatiotemporal trajectories of ERZ-level functionality loss under diverse rainfall scenarios. The framework does not depend on a specific hydrodynamic or damage model; any approach capable of producing coherent flood depth and impact estimates may be used. Notably, offline training does not require historical flood events and can use synthetic or design rainfall scenarios that adequately span local intensity–duration–frequency characteristics. While less realistic than observed data, these scenarios provide a practical alternative for regions with limited historical records.

Together, these data categories define the minimum information required to deploy CRAF in a new urban setting. The specific datasets used in the Fuzhou case study are summarized in Table 2.

### 3.2 Testbed Community and Data Instantiation

The urban area of Fuzhou, China, is used to demonstrate data preprocessing, offline training, and real-time deployment of the CRAF framework. As shown in Fig. 3a, a residential building inventory of 553 potentially impacted structures in the study area is classified into four dominant archetypes: high-rise, multi-story, detached villa, and overlay villa. Buildings are grouped into $M$=50 ERZs based on spatial proximity and flood-conditioning characteristics (Yang et al. 2021, Zhou et al. 2024), yielding response units consistent with emergency management practice (Fig. 3b).

For each ERZ, building archetype proportions are aggregated to form the ERZ-level attribute matrix $E$ (Fig. 3c). Meteorological forcing is derived from multiple historical typhoon events to cover a range of rainstorm intensities for offline training, yielding $N_R$ =752 rainstorm sequences of 24 h length; the demonstration event (i.e., Typhoon Haikui, 2023) is designated as the independent test set to avoid information leakage. Flood-conditioning variables include elevation-derived features and proximity indicators related to river networks and drainage outlets. These variables are initially computed at high spatial resolution and subsequently aggregated to the ERZ level to form the conditioning matrix $W$.

Table 2. Summary of data sources for the Fuzhou case study

| Data category | Indicator | Symbol | Resolution | Sources |
|---|---|---|---|---|
| ERZ-specific building attributes | Proportions of archetype | $E \in \mathbb{R}^{M \times d_e}$ | - | Amap (lbs.amap.com) |
| Meteorological variables | Hourly rainfall sequences | $R \in \mathbb{R}^{(H+F) \times N_R}$ | 1 hour | Rainfall sequences of 9 historical typhoon events* at 2122 meteorological stations from Fujian Provincial Flood Information Release System (https://slt.fujian.gov.cn/) |
| Flood conditioning variables | Digital Elevation Model (DEM) | $W \in \mathbb{R}^{M \times d_w}$ | 30m | Geospatial Data Cloud (www.gscloud.cn) |
| | Aspect | | 30m | Transformation from DEM data based on ArcGIS software |
| | Curvature | | 30m | |
| | Slope | | 30m | |
| | Topographic wetness index (TWI) | | 30m | |
| | Proximity to river network | | - | Visual interpretation from Google remote sensing imagery |
| | Proximity to drainage outlet | | - | Fuzhou Water Resources Bureau |
| Supervisory labels | ZFL trajectories | $Z \in \mathbb{R}^{(H+F) \times M}$ | Hourly per ERZ | Physic-based simulation (detailed in Section 3.3) |

* **Notes:** The 9 typhoon events—Soudelor (2015), Meranti (2016), Megi (2016), Nesat (2017), Haitang (2017), Lekima (2019), Lupit (2021), Doksuri (2023), and Haikui (2023)—provides 72-hour rainfall sequences (24 h pre-landfall to 48 h post-landfall). Sequences are segmented using a 24-hour sliding window with 12-hour stride, retaining only subsequences with accumulated rainfall exceeding 350 mm.

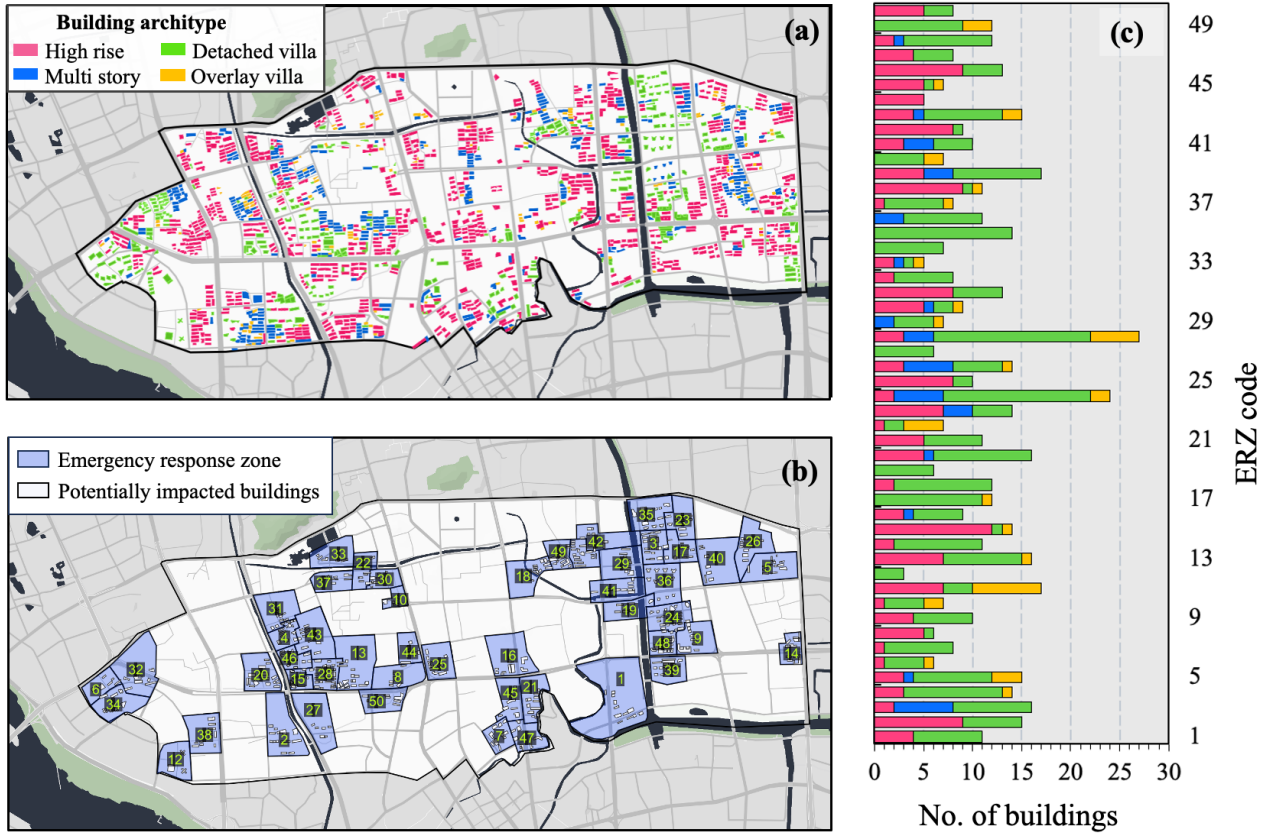

Fig. 3 (a) Building inventory in Fuzhou study area; (b) spatial distribution of ERZs; (c) archetype composition of each ERZ.

## 3.3 Physics-Based Simulation for Offline Supervisory Training

Physics-based simulations are employed to generate supervisory data for offline training of the SA and STF modules. In the testbed implementation, a two-stage simulation workflow is adopted (Fig. 4).

First, urban flood inundation is simulated using a coupled 1D–2D hydrodynamic framework that integrates the Storm Water Management Model (SWMM) with the Weighted Cellular Automata 2D model (WCA2D), resolving both drainage network flow and surface inundation under historical rainfall sequences (Tavakolifar et al., 2021; Zeng et al., 2022). Second, building functionality is assessed through a three-level fragility analysis, progressing from Needs-Related Components (NRCs) to individual dwelling units and finally to the entire residential building, thereby establishing depth-to-functionality mapping criteria. Simulated water depths are then translated into probabilistic building functionality states—considering physical damage, utility disruption and transportation accessibility—generating a seven-level functionality state (I–VII) probability vector for each building (Xie et al., 2025a). ERZ-level functionality is subsequently derived by aggregating building-level probabilities of non-functional states ($\mathcal{L}$={III, ..., VII}) within each ERZ, resulting in ZFL trajectories following the formulation in Section 2.

Repeating this process across all historical rainstorm scenarios yields an ensemble of spatiotemporally correlated ZFL trajectories, which provide structured and physically consistent supervisory labels for offline training of both the SA and STF modules. Importantly, within CRAF, physics-based simulation serves solely as a source of prior knowledge for model training and is not used for real-time forecasting during deployment.

**Step 1: Urban flood inundation modeling**

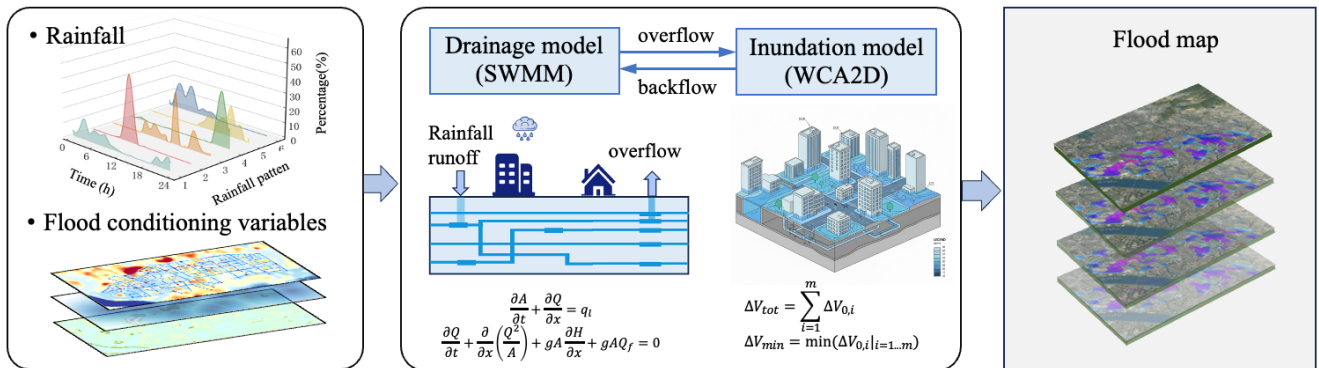

**Step 2: ERZ functionality loss assessment**

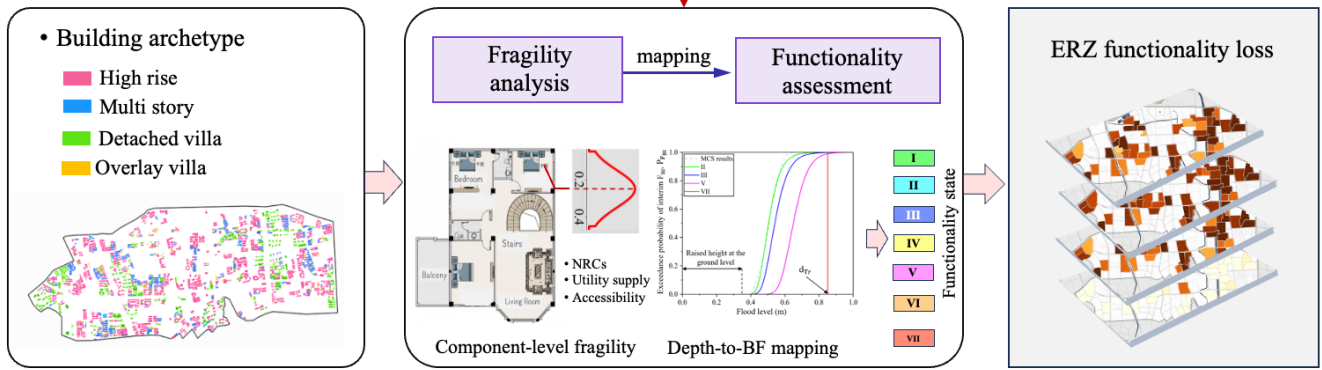

Fig. 4 Flowchart of the physics-based simulation workflow for generating ZFL trajectories.

## 4 DEVELOPMENT OF CRAF FRAMEWORK

This section details the technical implementation of CRAF, focusing on the architectures of its constituent modules and the associated learning strategy. Although CRAF operates as a unified observation–state–forecast process during deployment, its modules are designed under a strict separation between offline learning and online inference. The SA and STF modules are trained offline using physics-based flood and functionality simulations (Section 3.3) to learn persistent spatial and temporal impact dependencies, while real-time operation is limited to forward inference and sequential state updating. The following subsections describe each module and its role in the end-to-end system.

### 4.1 Crowdsourced Impact Monitoring (CIM): Observation Operator

The CIM module serves as the online observation operator of CRAF. Its role is to convert noisy, heterogeneous crowdsourced data into sparse but quality-controlled ERZ-level observations of ZFL.

As illustrated in Fig. 5, CIM is implemented as a modular, multimodal perception and inference framework, rather than a monolithic end-to-end network. First, flood-related content is collected from near–real-time social media and web platforms (including Weibo, TikTok, Tencent Docs, and various news portals) using keyword-driven crawlers (Table 3). Second, a relevance filtering step removes non-flood-related posts and retains only entries containing credible spatiotemporal and inundation cues. Textual relevance is determined using a fine-tuned BERT-based model (Kaur & Kaur, 2023), while content from other sources undergoes manual verification. Third, fine-grained flood attribute mining derives time, location, and water depth cues for each post. Specifically, location in text data is derived from check-ins, if available, or inferred via Point of Interest (POI)-based regular expression (regex) matching. Images and videos are manually interpreted to identify flooding scenarios and landmarks, which are then georeferenced by cross-checking with base maps or street-view data to derive coordinates. Flood depth cues in texts are also extracted via regex matching. Meanwhile, images and videos are processed through a physics-informed, rule-based mapping layer that translates qualitative inundation cues into quantitative flood-depth estimates using a hierarchical reference framework (Table 4). By combining location and flood depth, these point-based estimates are spatially interpolated via a constrained region-growing algorithm (Lin et al., 2023). Finally, water depths are mapped to ERZ-level ZFL using the same depth–functionality relationships adopted in the physics-based simulations (Section 3.3), ensuring consistency between offline supervisory labels and online observations.

To maintain reliability in the inference pipeline, the CIM module incorporates manual inspection and validation, correcting mislabeled or ambiguous entries and filtering out outliers or low-credibility reports before flood attribute extraction. By ensuring that only high-quality observations enter the pipeline, CIM avoids explicit error modeling, which would require large sample sizes and is impractical for sparse or unevenly distributed social media data, making manual quality control a key step for accurate ERZ-level functionality inference.

Table 3. Summary of flood-related keywords

| Category | Keywords |
| --- | --- |
| Related to rainfall | rainstorm (暴雨), heavy rain (大雨), |
| Related to typhoon | typhoon (台风), Haikui (海葵) |
| Related to flooding | flooding (洪涝), inundation (淹), waterlogging (积水), waist-deep (齐腰深), knee-deep (齐膝深). |

Table 4. Flood-depth hierarchical mapping strategy

| Reference object | Level | Part | Flooding depth estimation | Reference |
|---|---|---|---|---|
| Human body (average human height of 1.7m) | A | Ankle | 0.1 | (Chaudhary et al. 2020; Yan et al. 2024) |
| | B | Calf | 0.3 | |
| | C | Knee | 0.45 | |
| | D | Thigh | 0.64 | |
| | E | Waist | 0.85 | |
| | F | Chest | 1.28 | |
| | G | Neck | 1.49 | |
| Shared bicycle | A | Center of the wheel | 0.3 | Self-measured |
| | B | Top of the wheel | 0.5 | |
| | C | Saddle | 0.6 | |
| Car | A | Center of the tire | 0.33 | (Hao et al. 2022; Songchon et al. 2023; Yan et al. 2024) |
| | B | Top of the tire | 0.66 | |
| | C | Door handle | 0.8 | |

Notes: The flooding depth estimation represents statistical averages based on expert knowledge.

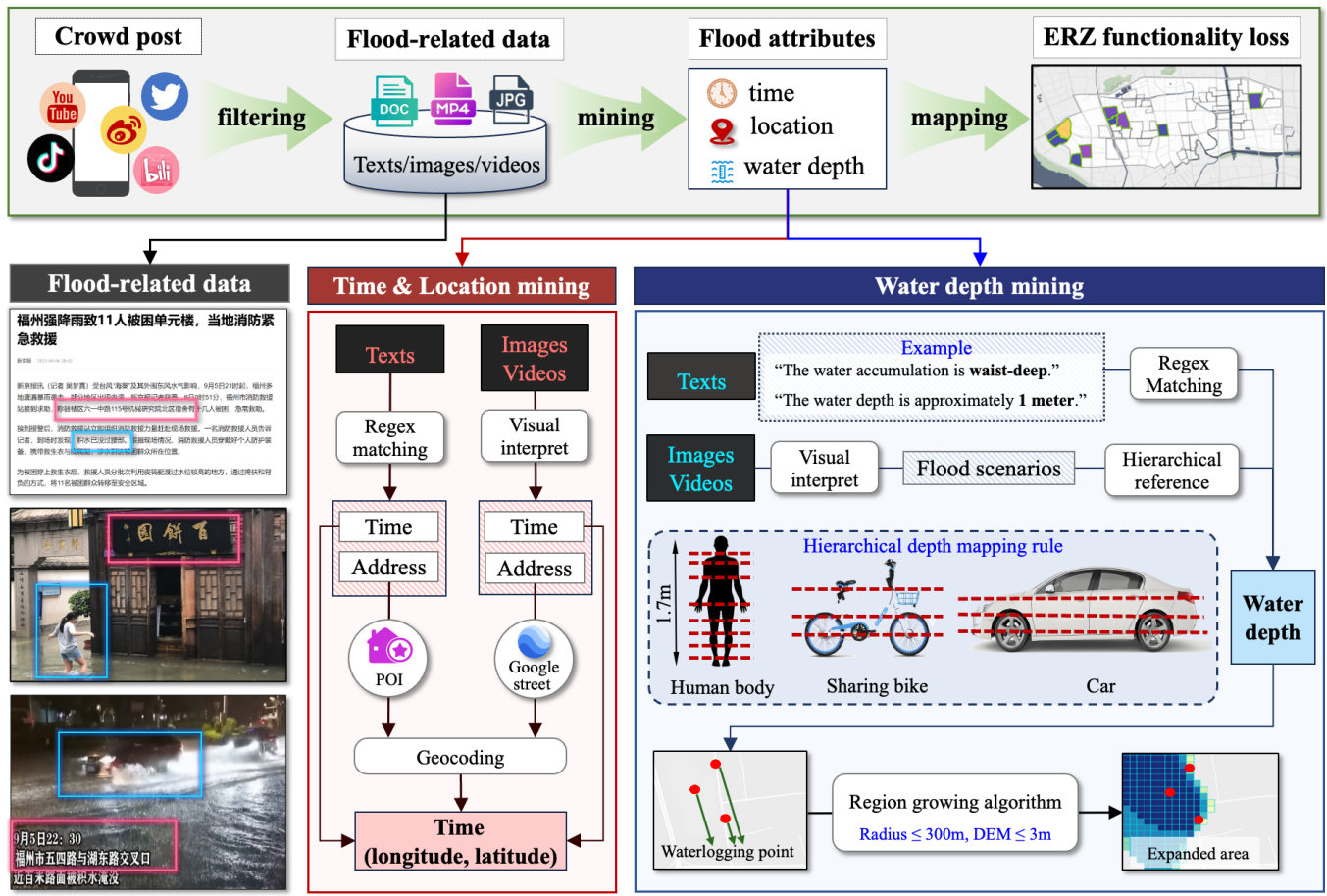

Fig. 5 CIM module architecture and its hierarchical flood attributes extraction mechanism.

## 4.2 Situational Awareness (SA): Physics-Informed Spatial State Completion

The SA module performs spatial impact-state completion, inferring a spatially complete and physically coherent ZFL snapshot from sparse ERZ-level observations. This step provides consistent initial conditions for downstream spatiotemporal forecasting and mitigates spatial artifacts arising from uneven observation coverage.

SA operates on an ERZ graph $G_1 = (V, A_1)$, where nodes represent ERZs and edges encode persistent inter-zone impact dependencies learned offline from physics-based simulations. Using simulated ZFL trajectories (Section 3.3), pairwise inter-ERZ dependencies are quantified and thresholded to construct a sparse adjacency structure that serves as a physics-informed structural prior during online inference.

SA is implemented as a multi-layer Graph Attention Network (GAT) performing masked node regression over the ERZ graph as illustrated in Fig. 6(a). Each ERZ node is associated with a feature vector comprising: (i) the observed ZFL value $z$ when available, otherwise an inverse-distance-weighted prior $\dot{z}$ ; (ii) aggregated building archetype proportions $E \in \mathbb{R}^{M \times d_e}$ ; and (iii) selected flood-conditioning variables $W^* \in \mathbb{R}^{M \times d_w^*}$ , including DEM, curvature, distances to rivers and drainage outlets. In the GAT, stacked multi-head attention layers propagate information via attention-weighted message passing, enabling inference at unobserved nodes conditioned on both local observations and physics-informed spatial dependencies. Observed nodes are clamped to reported values, while unobserved nodes are estimated through learned attention mechanisms.

Offline training uses physics-generated ZFL snapshots with synthetically masked observations to emulate realistic crowd-data sparsity (Fig. 6(b)). The training objective minimizes reconstruction error exclusively on masked nodes, promoting robust spatial inference under observation scarcity. Separate SA models are trained for different observation densities and reused unchanged during real-time deployment.

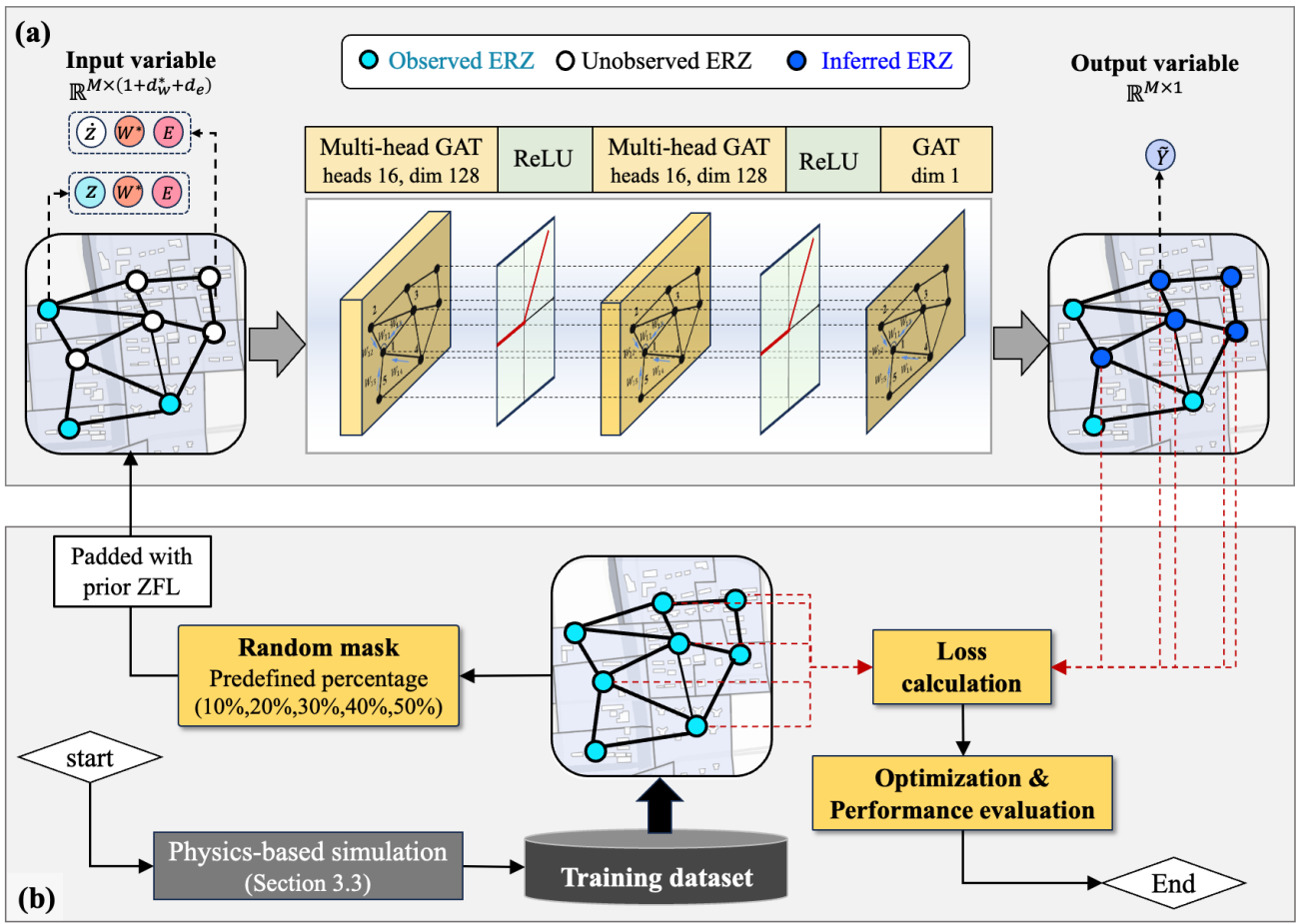

Fig. 6 SA module (a) architecture and (b) training strategy

## 4.3 Spatiotemporal Forecasting (STF): Rainfall-Conditioned State Propagation

The STF module performs rainfall-conditioned temporal state propagation, generating rolling multi-step forecasts of ERZ-level ZFL by advancing the completed impact state forward in time.

As shown in Fig. 7, STF couples rainfall forcing encoding with a spatiotemporal graph neural forecasting backbone. Given the rainfall sequence $R_{t-H+1:t+F}$, a temporal encoder extracts forcing features over the look-back and forecast horizon. These features are fused with the historical ZFL state sequence, consisting of SA-completed snapshots at observed times and model-imputed states at unobserved times. The fused representation is processed by stacked spatiotemporal convolution (ST-Conv) blocks, which combine temporal gated convolutions (e.g., GLU-based causal convolutions) with graph-based spatial aggregation to capture cross-ERZ propagation dynamics (Yu et al. 2018). Spatial coupling is defined on an ERZ graph $G_2 = (V, A_2)$, where $A_2$ encodes persistent inter-zone co-variation learned from physics-generated training trajectories. The output layer applies two temporal gated convolutions and a 2-D convolution to produce single-step forecasts.

STF is trained offline using physics-generated rainfall–impact pairs $(R, Z)$ under a sliding-window sampling strategy (Fig. 8). A 13-hour window with a 1-hour stride is applied to each 24-hour sequence, with zero-padding at the start. This yields 24 training pairs per sequence, for a total of 18,048 samples. The model is optimized for one-step-ahead prediction, while multi-step forecasts during both training and deployment are obtained via autoregressive rollout, consistent with real-time rolling-forecast operation.

Section 4 establishes CRAF as a modular, tightly integrated, physics-informed impact forecasting system. System performance is evaluated next from two complementary perspectives: controlled physics-based simulation experiments to establish baseline behavior and generalization (Section 5), followed by real-world application during an operational flood event to demonstrate practical effectiveness under uncertainty (Section 6).

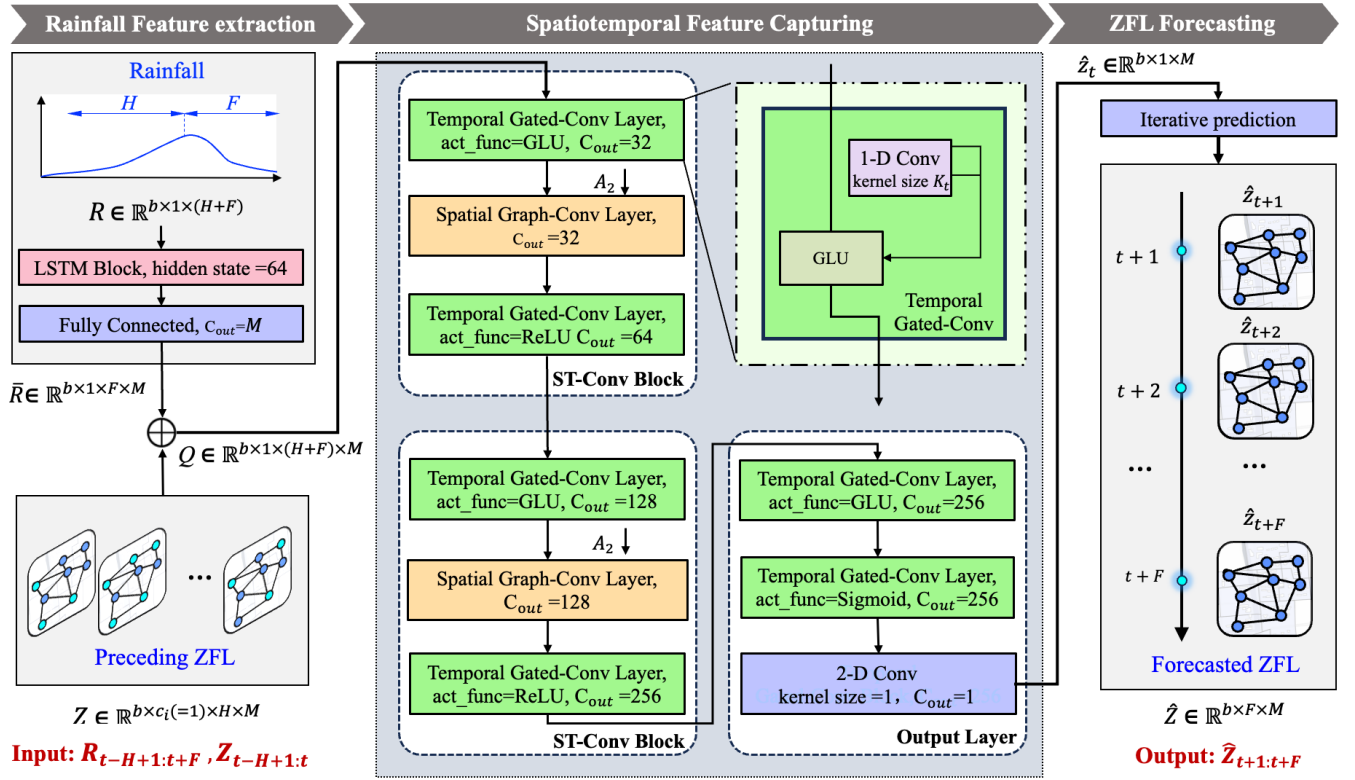

Fig. 7 STF module architecture.

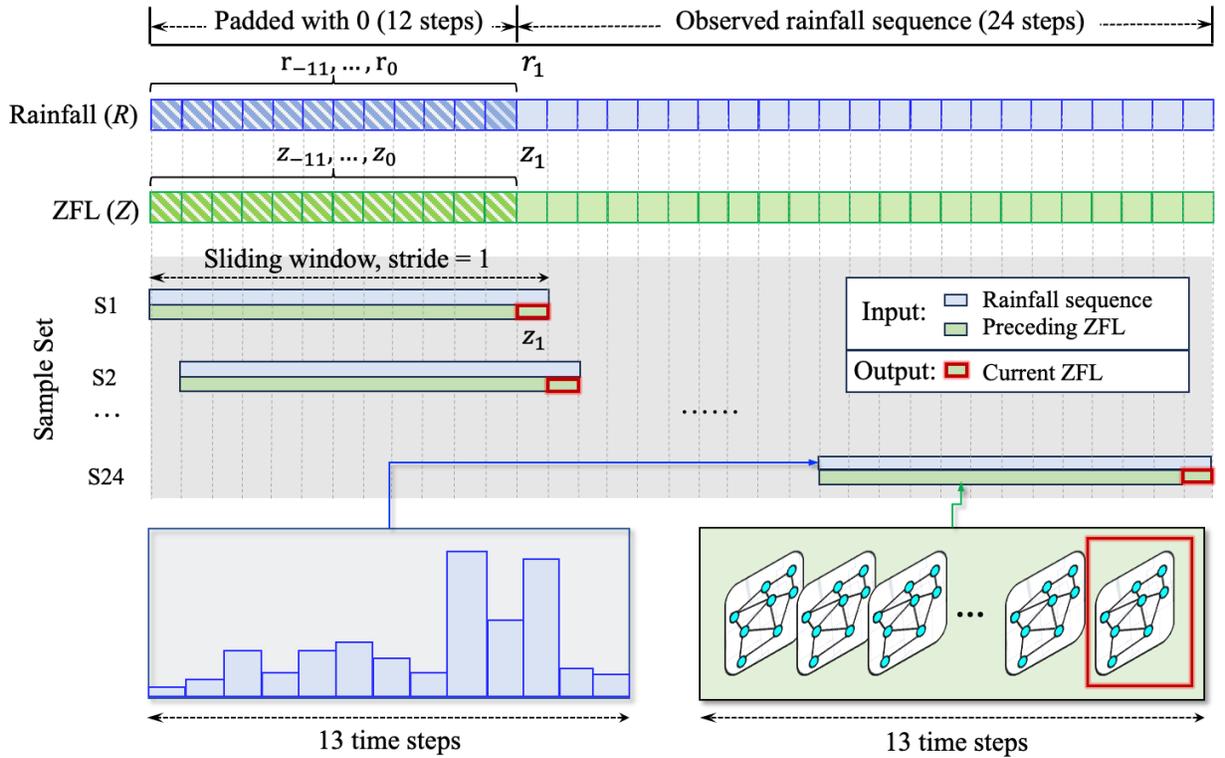

Fig. 8 Sliding-window training and autoregressive rollout strategy for STF.

## 5 RESULTS I - OFFLINE VERIFICATION USING PHYSICS-BASED SIMULATIONS

This section presents an offline verification of the CRAF framework using physics-based simulations. The objective is to assess the reliability, robustness, and cross-event generalization of the closed-loop forecasting process under controlled conditions, isolating the intrinsic behavior of the system prior to real-time deployment. By evaluating spatial state reconstruction and temporal propagation against known ground truth, these experiments establish whether the framework can provide stable and physically consistent forecasts independent of crowdsourced noise and operational uncertainties.

All numerical experiments are conducted using the physics-generated flood and building-functionality trajectories described in Section 3.3. Model training, validation, and testing rely exclusively on simulated data with known ground truth, enabling systematic assessment of spatial inference and temporal propagation behavior without confounding observational noise.

### 5.1 Verification of SA Module Under Sparse Observations

This experiment evaluates the ability of SA module to reconstruct spatially complete ZFL fields from sparse and uneven ERZ-level observations.

Physics-generated ZFL snapshots serve as ground truth. To emulate realistic observation scarcity, ERZ-level observations are synthetically masked at observation ratios of 10%, 20%, 30%, 40%, and 50%, with observed ERZs selected uniformly at random for each snapshot. Masked ERZs are treated as unobserved and inferred solely through the physics-trained situational prior. The physics-generated dataset designates Typhoon Haikui as the hold-out test set, with the remaining samples split into training and validation (7:3) for hyperparameter calibration. Performance is quantified using Mean Absolute Error (MAE) and Root Mean Squared Error (RMSE), computed exclusively over masked ERZs.

Table 5 summarizes reconstruction errors across training, validation, and testing under progressively sparse observations. Mean MAE ranges from 0.024 to 0.051 and RMSE from 0.048 to 0.093 across all observation ratios. Importantly, errors remain low even when only 10–20% of zones are observed, as also shown in Fig. 9, indicating that the learned spatial dependencies provide reliable inference under severe information scarcity. This behavior suggests that the physics-informed prior supplies a stable structural constraint, enabling consistent state estimation across storms rather than event-specific fitting. Such robustness is critical for operational settings where observations are incomplete and unevenly distributed.

Table 5. MAE and RMSE of the SA model under sparse ERZ-level observations

| Error metric | MAE | | | | | RMSE | | | | |
|---|---|---|---|---|---|---|---|---|---|---|
| Obs. ratio | 10% | 20% | 30% | 40% | 50% | 10% | 20% | 30% | 40% | 50% |
| Train set | 0.051 | 0.032 | 0.033 | 0.032 | 0.033 | 0.093 | 0.061 | 0.060 | 0.058 | 0.058 |
| Validation set | 0.042 | 0.027 | 0.025 | 0.024 | 0.025 | 0.083 | 0.055 | 0.048 | 0.049 | 0.048 |
| Test set (Haikui) | 0.044 | 0.030 | 0.030 | 0.030 | 0.03 | 0.087 | 0.062 | 0.060 | 0.059 | 0.058 |

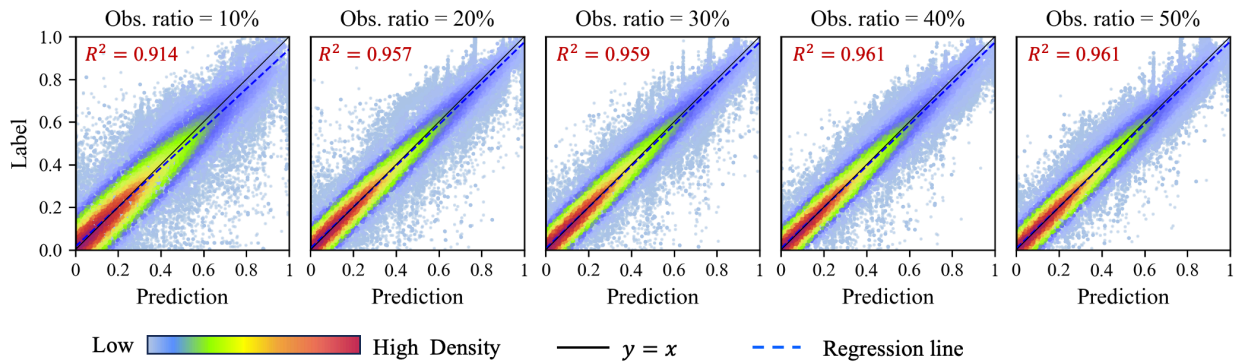

Fig. 9 SA model performance on the test set (Typhoon Haikui) across varying ERZ-level observations.

## 5.2 STF Module Temporal Stability and Forcing Consistency of Spatiotemporal Forecasting

This experiment examines the temporal behavior of the STF module under multi-step autoregressive rollout, to assess (i) the forecast stability over increasing lead times, and (ii) whether the forecast evolution is genuinely conditioned on meteorological forcing rather than dominated by autoregressive persistence.

Figs. 10 summarizes the performance of the STF module on the held-out Typhoon Haikui, presenting both temporal and spatial error patterns. Fig.10(a) shows the MAE evolution over a 24-hour iterative forecast window, indicating that errors increase smoothly during early lead times and stabilize at longer horizons. It also compares simulated and predicted ZFL values at multiple typical forecast horizons, showing that strong linear agreement is maintained across lead times without drift or mean collapse. Although Prediction dispersion increases gradually with lead time, as expected, no systematic bias amplification is observed. Fig.10(b) displays the corresponding spatial distribution of forecast errors across ERZs, with most zones maintaining low errors and a 24-h average below 1.52%.

Together, these results demonstrate that forecast errors grow gradually and remain bounded over extended lead times, indicating stable multi-step propagation without numerical drift or bias amplification. The absence of error escalation and the preservation of spatial structure suggest that the model maintains physically consistent dynamics during iterative rollout, a prerequisite for dependable rolling forecasts in real-time operations. From a reliability standpoint, this stability ensures that forecast quality degrades predictably rather than abruptly, supporting continuous situational awareness for emergency decision-making.

To verify that STF forecasts are genuinely conditioned on meteorological forcing, a rainfall-removal ablation (STF-NR) is conducted in which rainfall inputs are excluded while all other model components are held fixed. As shown in Table 6, removing rainfall forcing results in an order-of-magnitude degradation in both MAE and RMSE across all horizons and produces unstable error growth. This sharp deterioration confirms that forecast skill is governed primarily by physically meaningful rainfall–impact relationships rather than autoregressive persistence. Consequently, the STF module preserves causal consistency with external forcing, a necessary condition for reliable forecasting under evolving meteorological conditions.

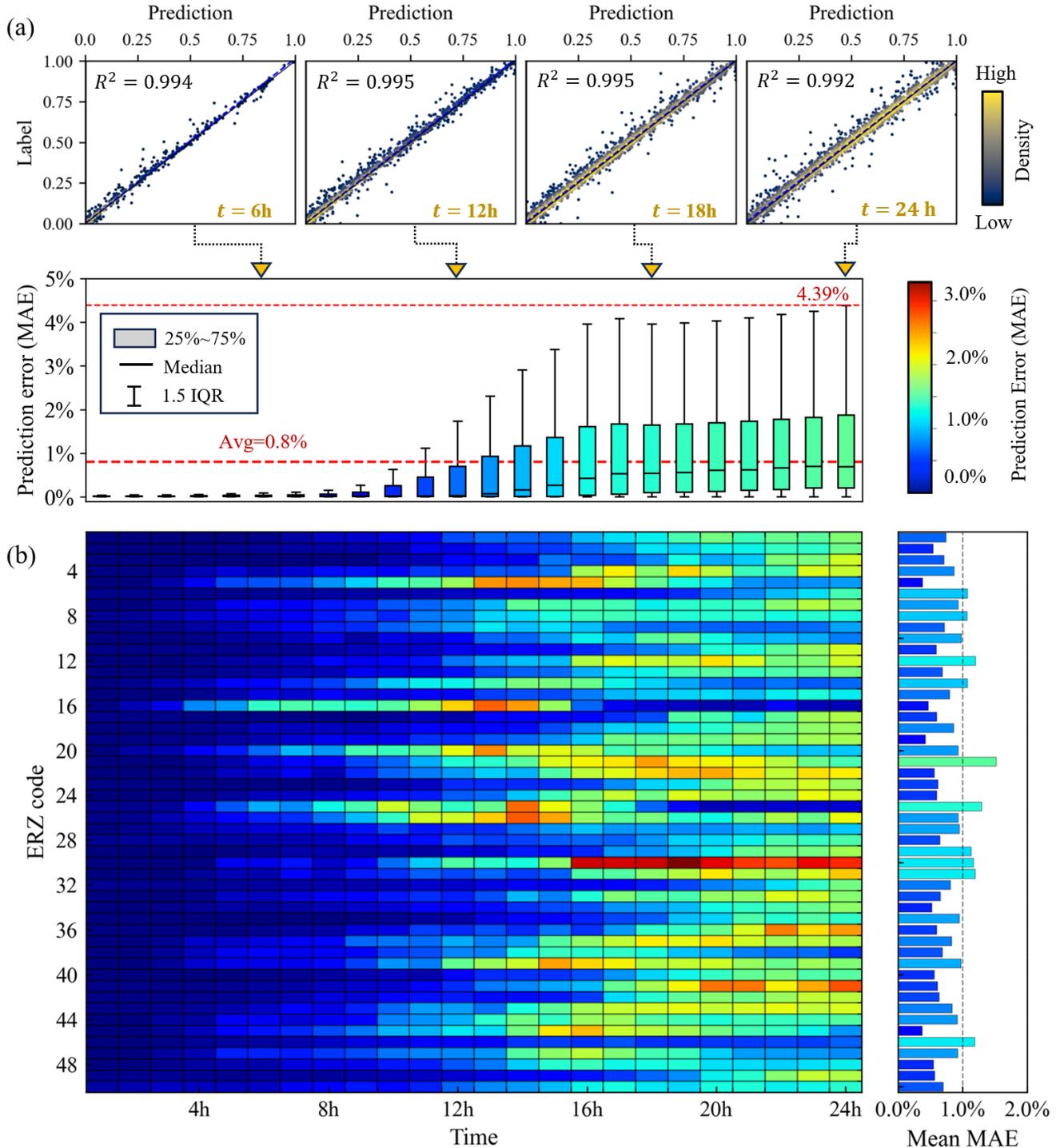

Fig. 10 (a) Temporal evolution of prediction errors (MAE) over a 24-hour iterative forecast window; (b) Spatial distribution of MAE across ERZs, with 24-hour average values.

Table 6. Effect of removing rainfall forcing on multi-step ZFL forecast accuracy under the LOSO protocol.

| Error metric | Model | Forecast horizons (hours ahead) | | | |
|---|---|---|---|---|---|
| | | +6 h | +12 h | +18 h | +24 h |
| MAE | STF | **0.0024** | **0.0077** | **0.0131** | **0.0155** |
| | STF-NR | 0.3712 | 0.5245 | 0.4095 | 0.2661 |
| RMSE | STF | **0.0102** | **0.0191** | **0.0247** | **0.0293** |
| | STF-NR | 0.4804 | 0.6031 | 0.5138 | 0.3810 |

Overall, the offline verification demonstrates that CRAF provides stable, physically consistent, and transferable spatial–temporal forecasting behavior across storms. The framework maintains reliable state reconstruction under sparse observations and stable multi-step propagation under dynamic forcing, establishing a dependable foundation for subsequent real-time, closed-loop deployment. Section 6 extends this evaluation to a closed-loop setting, examining its performance under real-world operational constraints.

## 6 RESULTS II- ONLINE DEMONSTRATION UNDER REAL-WORLD CONDITIONS

### 6.1 Demonstration Setup and Operational Context

This section evaluates the operational behavior of the CRAF framework during a real flood event, assessing end-to-end system reliability under realistic conditions characterized by uncertain rainfall forcing, sparse observations, and time-critical decision constraints.

The demonstration is conducted using the severe rainstorm associated with Typhoon Haikui (5–6 September 2023) in Fuzhou, China. This event is held out entirely from training and validation and serves as an independent operational test. Its rapid rainfall intensification, substantial numerical weather prediction bias, and uneven availability of real-time observations create a representative stress scenario rather than a best-case setting.

Evaluation is framed from a decision-support perspective. Emergency managers operate on hourly update cycles and must issue evacuation advisories, activate shelters, and allocate resources within short lead times. Accordingly, the analysis emphasizes (i) the reliability of situational awareness under sparse crowdsourced evidence, (ii) the ability of rolling forecasts to correct rainfall-driven bias through impact-state updating, and (iii) forecast stability within short horizons (1–3 h) that align with practical decision windows. Results are interpreted in terms of operational behavior and system robustness rather than location-specific performance.

### 6.2 Rainfall Forecast Uncertainty and Impact Forecasting Challenges

Accurate impact forecasting during urban flood events is fundamentally constrained by the quality of rainfall forcing. Fig.11(a) compares observed hourly rainfall during Typhoon Haikui with the operational numerical weather prediction used for forecasting. The forecast substantially underestimates both rainfall intensity and cumulative accumulation, producing 162.6 mm compared to the observed 447.5 mm over the event period.

When forecasts underestimate rainfall, hydrologic and hydraulic models propagate this bias downstream, leading to systematic underestimation of flood extent and associated functionality loss. Consequently, rainfall-driven impact forecasts alone may fail to capture both the timing and magnitude of escalating disruptions during the most critical response window.

Rather than attempting to correct meteorological inputs—which is rarely feasible within operational time constraints—CRAF performs correction directly in the impact space by assimilating real-time human-sensed observations (Fig. 11(b)). This strategy shifts uncertainty management from upstream hazard estimates to downstream consequences, where decisions are ultimately made. The following analyses demonstrate how this impact-space updating stabilizes situational awareness and improves forecast reliability under severe rainfall forcing errors.

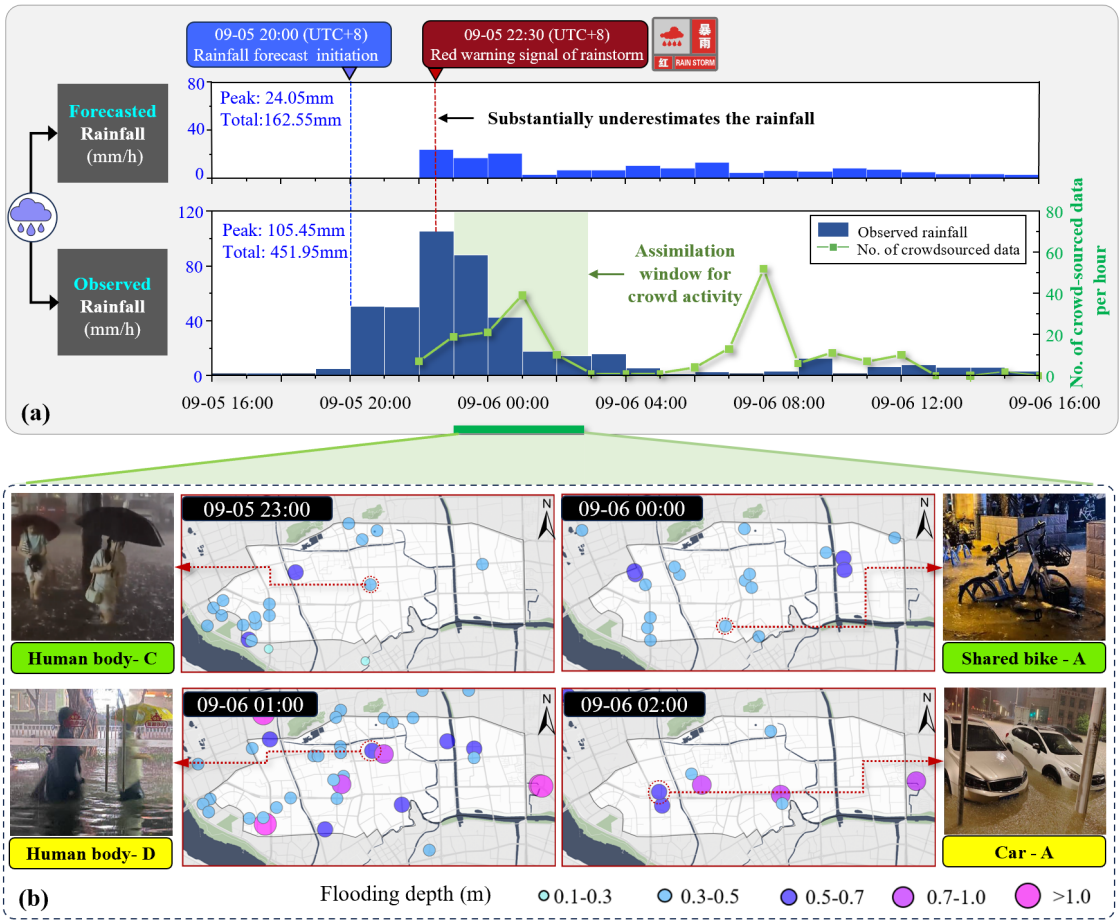

Fig. 11 Rainfall and waterlogging scenarios: (a) temporal comparison of forecasted and observed hourly rainfall and crowd activity; (b) representative waterlogging observations at different time steps.

## 6.3 Situational Awareness Enhancement Under Sparse Crowd Observations

Accurate short-horizon impact forecasting hinges on the quality of the *current* impact-state estimate. During Typhoon Haikui, this requirement is especially acute: rainfall forcing is severely underpredicted, while crowdsourced observations—though informative—are sparse, uneven, and temporally clustered. As shown in Fig. 12(a), ERZ-level observation coverage during the first major crowdsourcing peak (22:00 on September 5 to 02:00 on September 6) spans only 16%–42%, leaving large portions of the urban system unobserved at precisely the moment when impacts are escalating.

If initialized directly from biased rainfall forecasts and raw sparse observations, any downstream forecasting—regardless of model sophistication—would be structurally misaligned with reality. The SA module addresses this failure mode by explicitly performing *current impact-state alignment* prior to forecasting. Leveraging a physics-trained spatial prior, SA propagates information from observed ERZs (shaded) to unobserved zones (unshaded), reconstructing a spatially complete and physically coherent ZFL field that reflects system-level impact conditions rather than isolated reports. As illustrated in Fig. 12(b), inferred ZFL patterns evolve smoothly across space, capturing coherent escalation dynamics instead of fragmented observational artifacts.

Importantly, SA does not merely "fill gaps" in observations. By encoding inter-ERZ impact co-variation learned from physics-based simulations, it constrains inference to physically plausible spatial dependencies, preventing underestimation under sparse coverage. In doing so, SA transforms heterogeneous crowdsourced evidence into a *decision-grade impact state* that reconciles human sensing with physical process constraints.

From an operational standpoint, this alignment step is foundational. Without spatial completion, downstream forecasts would inherit compounded bias from rainfall underestimation and incomplete observations, rapidly eroding decision confidence. By reconstructing a calibrated, system-consistent impact state prior to propagation, SA provides a reliable initialization that anchors subsequent forecasts to evolving ground truth. This capability transforms sparse and heterogeneous crowd reports into actionable situational awareness and is essential for dependable real-time impact forecasting under precisely the conditions where rainfall-driven approaches are most vulnerable.

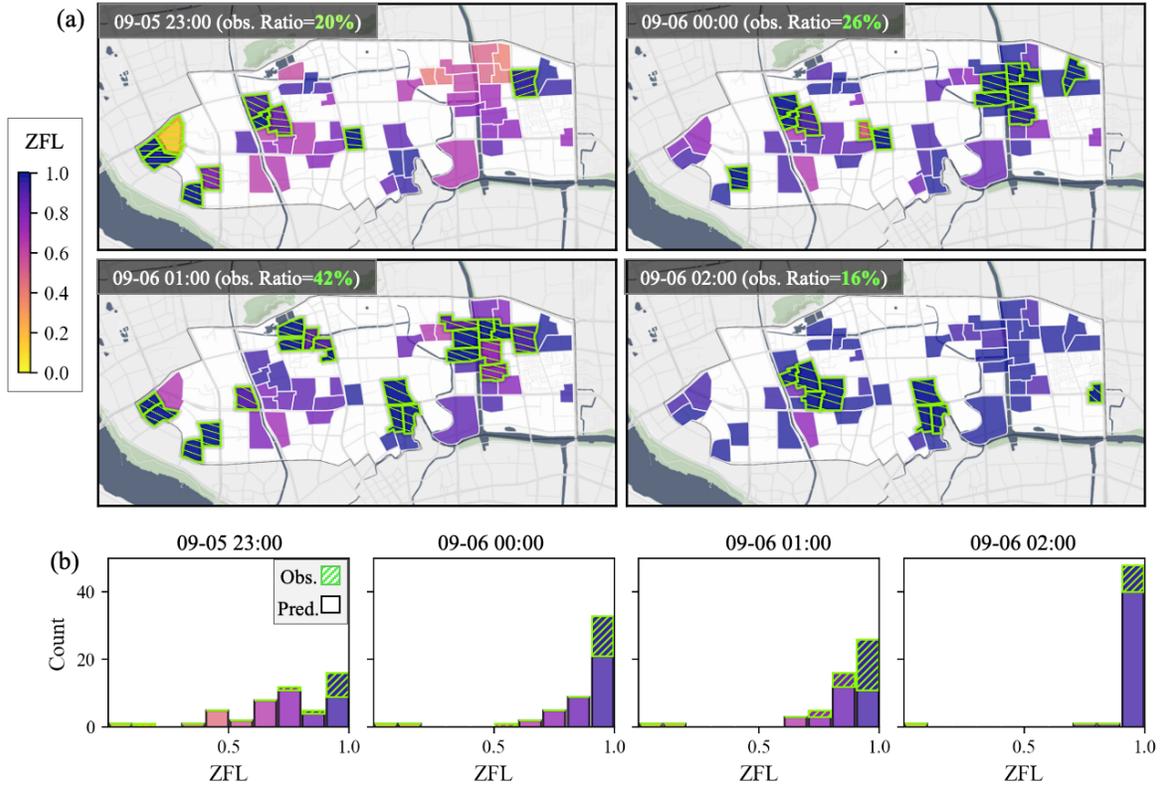

Fig. 12 Spatiotemporal distribution of ZFL across four time slices.

### 6.4 Rolling Forecast Updating and Error Reduction During Critical Decision Windows

Building on the calibrated impact states produced by the SA module, rolling multi-step forecasts are evaluated under four clearly defined information regimes, with the first two serving as baseline scenarios:

1. **STF-OL-FR** (Open-Loop, Fixed Rainfall-driven Forecasting): represents a one-time forward propagation of the spatiotemporal forecast over the entire horizon using fixed rainfall forcing, without any rolling updates or assimilation of impact observations;

2. **STF-OL-UR** (Open-Loop, Updated Rainfall-driven Forecasting): at each cycle the spatiotemporal forecasting model is driven by updated rainfall (from numerical weather predictions and ground meteorological station observations), but without assimilating impact observations;

3. **CRAF** (Closed-Loop CIM+SA+STF Forecasting): STF is reinitialized at each update cycle using SA-completed impact states that assimilate crowdsourced observations, in combination with updated rainfall forcing; and

4. **Observed ZFL**: the reference impact state derived from aggregated crowd evidence, serving exclusively as the evaluation benchmark.

Figs. 13 compare rolling forecasts produced by STF-OL-FR and CRAF across successive update cycles during Typhoon Haikui. Across successive update cycles, open-loop forecasts (STF-OL-FR) systematically underestimate both the magnitude and timing of functionality loss due to rainfall forcing bias, producing trajectories that are internally consistent yet operationally misleading. In contrast, closed-loop CRAF forecasts remain closely coupled to the observed impact evolution. By assimilating crowdsourced evidence before each rollout, CRAF repeatedly realigns the system state with current conditions, preventing error accumulation and maintaining forecast credibility. This behavior yields consistent 1–3 h-ahead error reductions of 84.4–95.1% relative to the fixed-forcing baseline, demonstrating stable closed-loop dynamics rather than isolated corrections.

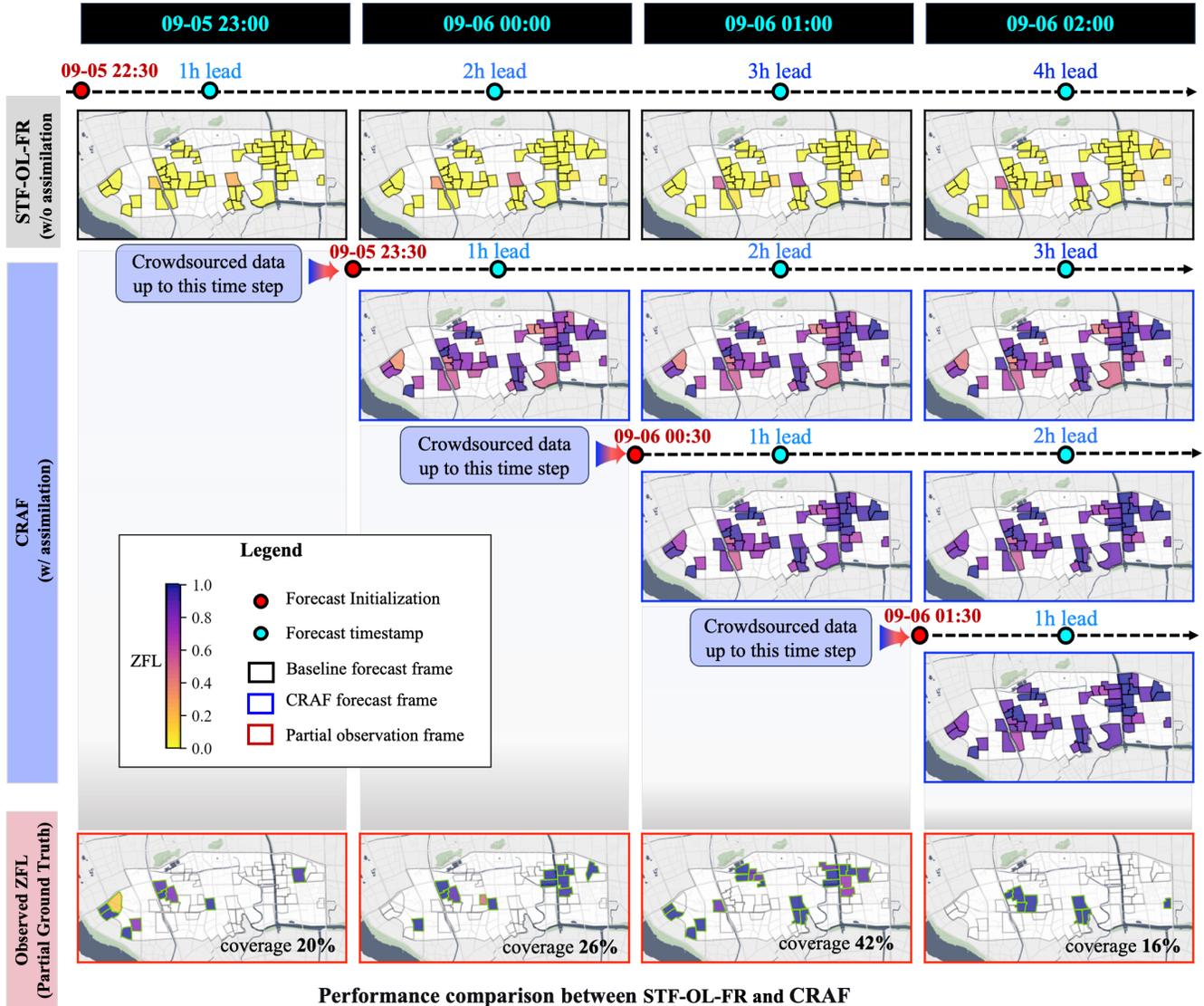

| Forecast Cycle | | 09-05 23:00 | 09-06 00:00 | 09-06 01:00 | 09-06 02:00 |
|---|---|---|---|---|---|
| | | ZFL (% improvement relative to baseline forecast) | | | |
| STF-OL-FR (fixed rainfall-driven forecast) | | 0.001 | 0.004 | 0.059 | 0.101 |
| CRAF (w/ assimilation) | assimilated at 09-05 23:30 | - | 0.805 (↓84.41%) | 0.835 (↓87.39%) | 0.867 (↓85.3%) |
| | assimilated at 09-06 00:30 | - | - | 0.903 (↓95.05%) | 0.899 (↓88.86%) |
| | assimilated at 09-06 01:30 | - | - | - | 0.916 (↓90.76%) |
| Observed ZFL (ground truth) (mean±std) | | 0.866 ± 0.257 | 0.953 ± 0.132 | 0.947 ± 0.085 | 0.999 ± 0.001 |

Fig. 13 Comparison of dynamic forecasting performance between STF-OL-FR baseline (fixed rainfall-driven one-time forecasting without crowd assimilation) and CRAF models across multiple initialization time steps.

Fig. 14 presents prediction error (MAE) comparisons between STF-OL-UR and CRAF. When rainfall inputs are updated but impact observations are not assimilated with STF-OL-UR, substantial errors persist. Even under identical rainfall forcing, CRAF reduces short-horizon MAE by 72.8–79.6%, and the error distribution becomes markedly more concentrated. This contraction reflects not only improved accuracy but also enhanced stability and predictability—properties that directly translate to greater operational trust. Together, these results show that reliability gains arise primarily from dynamic state alignment rather than rainfall refinement alone.

Given that time-sensitive emergency actions (e.g., evacuation advisories, shelter activation, and resource deployment) are typically initiated within narrow 1–3 h decision windows, this dual merit of CRAF carries decisive practical weight. In stark contrast, open-loop rainfall-driven forecasts can deviate rapidly under forcing uncertainty, undermining trust and operational usefulness. By closing the loop between sensing and forecasting, CRAF continuously realigns predictions with evolving ground truth at low latency. This ability to stabilize forecasts under uncertainty—rather than merely propagate hazards—defines the practical value of closed-loop impact forecasting and distinguishes CRAF from conventional open-loop approaches.

During the Haikui event, CRAF completed a full update cycle—including crowd ingestion, situational inference, and multi-step forecasting—in approximately 10 minutes per update on a standard GPU. This latency comfortably supports hourly or sub-hourly rolling updates required in operational emergency response. To ensure reliability, crowdsourced data assimilation is restricted to the event escalation phase, as post-peak reports may underestimate residual functionality loss. Together, these design choices reflect practical deployment constraints and ensures timely yet stable real-time operation.

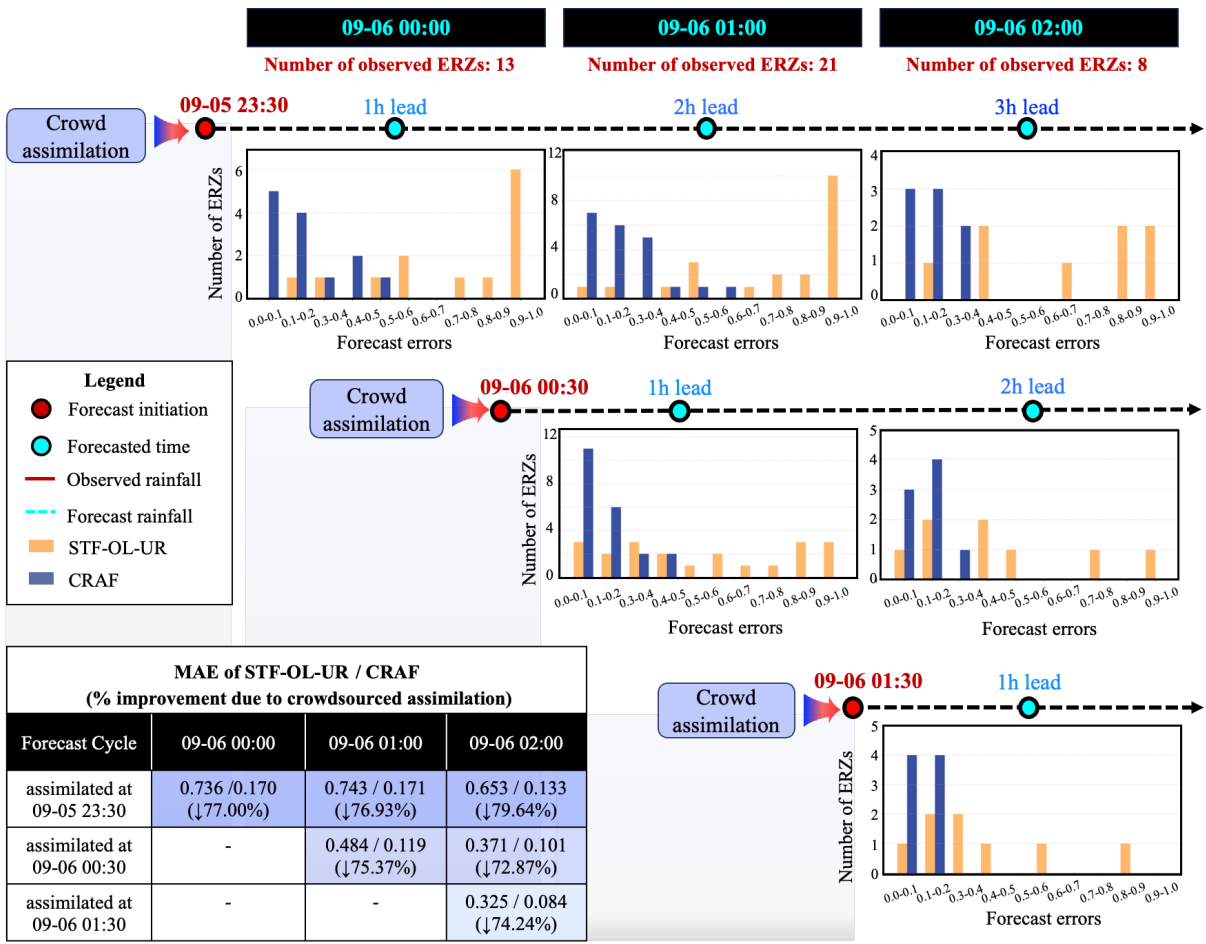

Fig. 14 Prediction error comparisons between STF-OL-UR and CRAF.

## 7 DISCUSSION AND CONCLUSIONS

This study introduced CRAF, a physics-informed, closed-loop impact forecasting framework designed to operate under two persistent challenges in real-world flood response: uncertain hazard forcing and sparse, heterogeneous observations. Through controlled offline verification and an operational deployment, the results demonstrate that reliable short-horizon impact forecasting depends less on refining hazard inputs than on continuously aligning the evolving impact state with real-world evidence. This finding reframes impact forecasting as a state estimation problem rather than solely a hazard prediction problem.

A central insight is that impact-state initialization dominates forecast skill under forcing uncertainty. When rainfall forecasts are biased, open-loop impact propagation rapidly diverges from reality, even if internally consistent. CRAF mitigates this failure mode by assimilating sparse crowdsourced observations through a physics-trained situational prior, producing spatially coherent impact states before forecasting. This alignment step is foundational rather than auxiliary: without it, downstream forecasts remain structurally misaligned regardless of model sophistication. By explicitly closing the loop between sensing, situational inference, and forecasting, CRAF departs from conventional open-loop hazard-driven approaches and delivers stable gains within decision-relevant horizons. In the demonstrated event, CRAF reduced 1–3 hour-ahead impact forecast errors by approximately 84.4%-95.1% relative to the fixed rainfall-driven baseline and 72.8%–79.6% relative to the updated rainfall-driven baseline, while maintaining 10-minute update latency suitable for real-time operation.

The robustness of the framework stems from a deliberate separation between offline physics-supervised learning and online inference and updating. Physics-based simulations are used offline to learn transferable spatial and temporal dependencies, while real-time operation is restricted to state updating and forward propagation. This design avoids instability associated with online retraining while enabling rapid, reliable updates under operational constraints. By prioritizing situational alignment over hazard perfection, CRAF offers a robust foundation for impact-based early warning systems and operational digital twins that must remain coupled to the real world under uncertainty.

Several limitations warrant acknowledgment. The effectiveness of situational correction depends on the availability of timely observational evidence during event escalation, and the physics-based simulations used for offline training reflect region- and system-specific modeling assumptions. Moreover, actual impacts inferred from CIM and SA remain reliable and independent of rainfall forecasts, but STF's forecasted impacts are influenced by rainfall inputs, with large deviations potentially accumulating errors in longer-term rolling forecasts. Future work will extend the framework across additional hazards, regions, and sensing modalities, explore adaptive weighting of observational uncertainty, and investigate strategies to mitigate rainfall-dependent forecast errors, such as adaptive parameter updating or inverse estimation of rainfall forcing.

## ACKNOWLEDGEMENT

This project is supported by Zhejiang Department of Science & Technology through Grant No. 2024C03255.